\documentclass{article} 
\usepackage{MLDD_workshop_2023,times}

\usepackage{amsmath,amsfonts,bm}

\def\eqref#1{equation~\ref{#1}}

\def\1{\bm{1}}

\DeclareMathAlphabet{\mathsfit}{\encodingdefault}{\sfdefault}{m}{sl}
\SetMathAlphabet{\mathsfit}{bold}{\encodingdefault}{\sfdefault}{bx}{n}

\newcommand{\KL}{D_{\mathrm{KL}}}

\usepackage[hidelinks]{hyperref}
\usepackage{url}

\title{EigenFold: Generative Protein Structure \\Prediction with Diffusion Models}

\author{Bowen Jing,\thanks{Correspondence to \texttt{bjing@mit.edu}}\;\;\textsuperscript{1} Ezra Erives,\textsuperscript{1} Peter Pao-Huang,\textsuperscript{3} Gabriele Corso,\textsuperscript{1}\\
\textbf{Bonnie Berger,\textsuperscript{1\,2} \textbf{Tommi Jaakkola}\textsuperscript{1} } \\
\textsuperscript{1}CSAIL, Massachusetts Institute of Technology \\
\textsuperscript{2}Dept. of Mathematics, Massachusetts Institute of Technology \\
\textsuperscript{3}Dept. of Computer Science, University of Illinois\\
}

\iclrfinalcopy 
\usepackage{lipsum,todonotes,amsmath, booktabs, caption, subcaption}

\newcommand{\bfx}{\mathbf{x}}
\newcommand{\bfy}{\mathbf{y}}

\newcommand{\bfs}{\mathbf{s}}
\newcommand{\bfP}{\mathbf{P}}

\newcommand{\bfH}{\mathbf{H}}
\newcommand{\bfw}{\mathbf{w}}
\DeclareMathOperator{\diag}{diag}
\newcommand{\bfLambda}{\boldsymbol{\Lambda}}
\begin{document}
\maketitle

\begin{abstract}
Protein structure prediction has reached revolutionary levels of accuracy on single structures, yet distributional modeling paradigms are needed to capture the conformational ensembles and flexibility that underlie biological function. Towards this goal, we develop \textsc{EigenFold}, a diffusion generative modeling framework for sampling a distribution of structures from a given protein sequence. We define a diffusion process that models the structure as a system of harmonic oscillators and which naturally induces a cascading-resolution generative process along the eigenmodes of the system. On recent CAMEO targets, \textsc{EigenFold} achieves a median TMScore of 0.84, while providing a more comprehensive picture of model uncertainty via the ensemble of sampled structures relative to existing methods. We then assess \textsc{EigenFold}'s ability to model and predict conformational heterogeneity for fold-switching proteins and ligand-induced conformational change. Code is available at \url{https://github.com/bjing2016/EigenFold}.
\end{abstract}

\section{Introduction}

The development of accurate methods for protein structure prediction such as AlphaFold2 \citep{jumper2021highly} has revolutionized \emph{in silico} understanding of protein structure and function. However, while such methods are designed to model static experimental structures from crystallography or cryo-EM, proteins \emph{in vivo} adopt dynamic structural ensembles featuring conformational flexibility, change, and even disorder to effect their biological functions \citep{teague2003implications, wright2015intrinsically}. These aspects  represent the next frontier towards a more complete understanding  of protein structure and function \citep{lane2023protein}. Accordingly, there is increasing need for \emph{generative} models for protein structure prediction that can produce a distribution of conformations for a single protein sequence.

Meanwhile, generative modeling in molecular machine learning has undergone a renaissance driven by the paradigm of \emph{diffusion models} \citep{sohl2015deep, song2021score}. When applied to problems such as protein design \citep{watson2022broadly},  molecular docking \citep{corso2022diffdock}, and ligand design \citep{schneuing2022structure}, such models have displayed impressive distributional modeling. These capabilities make diffusion models compelling tools for understanding protein structural ensembles given a fixed sequence, but they have yet to be explored for this purpose.

\begin{figure}
    \includegraphics[width=\textwidth]{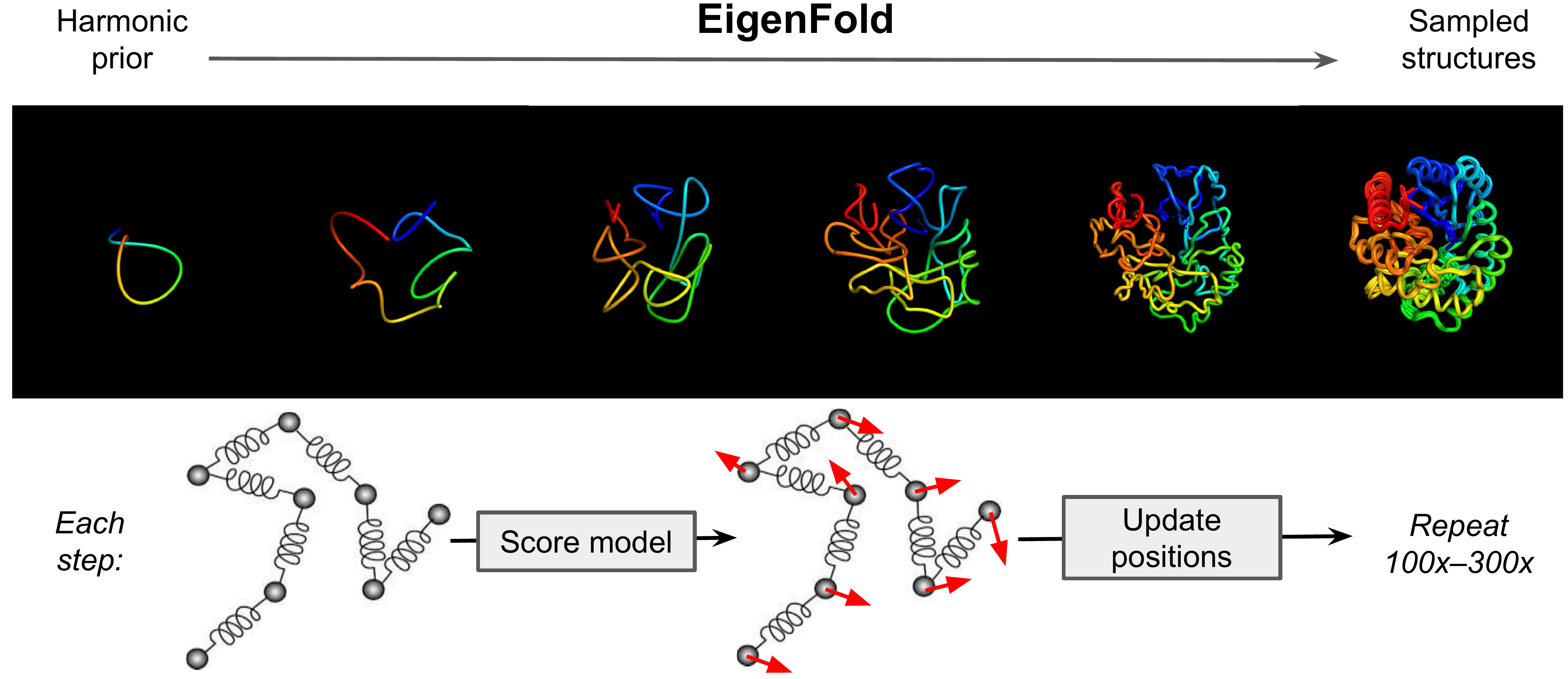}
    \caption{Overview of generative structure prediction with \textsc{EigenFold}. Samples from the harmonic diffusion prior (left) are gradually transformed into complete structures via a cascading-resolution generative process. One complete trajectory and five final sampled structures for CAMEO target \texttt{7dz2.C} are shown here. In each denoising step, the score model predicts update "forces," which are combined with harmonic constraints to update the residue positions. Depending on protein size, this procedure is repeated 100 to 300 times.}
    \label{fig:overview}

\end{figure}

To bridge this gap, we develop \textsc{EigenFold}, the first diffusion generative modeling framework for protein structure (and structural ensemble) prediction. To do so, we formulate a novel diffusion process---\emph{harmonic diffusion}---that models the molecule as a system of harmonic oscillators. The structure is projected onto the eigenmodes (or normal modes) of the system during the forward diffusion, such that the corresponding reverse diffusion can be viewed as a cascading-resolution generative process---first sampling the rough global structure before refining local details. This enables \textsc{EigenFold} to accurately sample protein structures with as few as $100$ inference steps. 

The \textsc{EigenFold} framework can be used in isolation or in conjunction with pretrained embeddings from existing structure prediction models. In this work, we train \textsc{EigenFold} using edge and node embeddings from {OmegaFold} \citep{wu2022high}, in effect transforming the latter into a generative model. When trained on the PDB and evaluated on CAMEO benchmarks, samples from this model are comparable with existing methods such as {RoseTTAFold} \citep{baek2021accurate} in terms of single-structure accuracy. However, unlike existing methods, \textsc{EigenFold} provides a \emph{distribution} of structures rather than scalar predicted errors, providing more insights into model uncertainty. In particular, we show that the \emph{variation} among the sampled structures is highly indicative of the model \emph{error} relative to the ground truth, for many metrics of model accuracy.

We then benchmark \textsc{EigenFold}'s ability to model conformational change and flexibility using two datasets: one of fold-switching proteins \citep{chakravarty2022alphafold2} and one of conformational changes associated with binding \citep{saldano2022impact}. The analysis yields mixed results, in which properties of \textsc{EigenFold} sampled structures are moderately correlated with properties of the ground-truth conformation but do not predict them to high accuracy. While not quite bridging the gap between single-structure prediction and structural ensemble prediction, these results and methodology lay the foundation for many possible directions for improvement in future work.

\section{Background and Related Work}

\textbf{Protein structure prediction.} The problem of predicting an experimental-level protein structure from sequence is widely considered to have been solved by AlphaFold2 \citep{jumper2021highly} in CASP 14. Since then, alternative models such as RoseTTAFold \citep{baek2021accurate}, ESMFold \citep{Lin2022ESM2}, and OmegaFold \citep{wu2022high} have replicated or approached similar levels of performance. All are developed and trained as \emph{deterministic} maps from input (sequence or MSA) to output (structure), making them suboptimal for modeling structural ensembles \citep{lane2023protein, chakravarty2022alphafold2, saldano2022impact}. MSA subsampling and clustering (i.e., varying the input) in conjunction with AlphaFold2 has recently been shown to reveal alternate conformations \citep{wayment2022prediction, stein2022speach_af, del2022sampling}, but the generality and reliability of these techniques remains unclear.

\textbf{Diffusion models} learn an iterative, stochastic generative process that transforms samples from a simple prior to the data distribution. This generative process is trained to be the reverse of a \emph{forward} diffusion transforming the data to the prior \citep{sohl2015deep, ho2020denoising, song2021score}. To obtain the generative process from the forward process, it is necessary and sufficient to learn a \emph{score model} to approximate $\nabla_\bfx \log p_t(\bfx)$ for all values of diffusion time $t$; we refer to \citet{song2021score} for a more comprehensive methodological overview. 

The flexible formulation and strong performance of diffusion models have made them increasingly popular in generative machine learning. While such models have traditionally used isotropic Gaussian noise as the forward process, applications for molecule structure have increasingly featured non-isotropic or non-Euclidean processes that exploit the reduced degrees of freedom and chemical priors in molecular structure \citep{jing2022torsional, ingraham2022illuminating}. Our work proceeds in a similar spirit and seeks to define a suitable diffusion process for protein structure prediction.

As generative models, diffusion models have been natural choices for inverse (design) problems. However, they have also been applied productively to forward problems such as ligand-protein docking \citep{corso2022diffdock} and molecular simulation \citep{wu2022score}. In particular, diffusion models operating over internal coordinates hold state-of-the-art performance on small-molecule \emph{conformer generation} \citep{jing2022torsional}. Generative protein structure prediction can be regarded as the macromolecular analogue of conformer generation; however, the significantly larger molecular graphs in protein structure call for different considerations and formulations.

\textbf{Protein structure diffusion.}
Several works have fruitfully applied diffusion modeling to the broadly defined task of \emph{protein structure design}. Early works formulated variations of isotropic Euclidean diffusion of residue coordinates \citep{anand2022protein, trippe2022diffusion} or backbone dihedral angles \citep{wu2022protein}. Later works demonstrated impressive experimental results \citep{watson2022broadly} and programmability \citep{ingraham2022illuminating}. On the other hand, there have been significantly fewer diffusion models developed for \emph{forward} problems involving protein structures (i.e., where the protein sequence is known) \citep{qiao2022dynamic, nakata2022end}. Both of these address the task of flexible protein-ligand docking, but have been limited by a dependence on contact maps and by protein size, respectively.

\section{Method}
\subsection{Harmonic Diffusion}
Consider a structure graph $G = (\mathcal{V}, \mathcal{E})$ embedded in 3D space with coordinates $\mathbf{x} \in \mathbb{R}^{3n}$, where $n = |\mathcal{V}|$. When $G$ represents a protein with a specific sequence, the \emph{generative protein structure prediction} problem can be framed as learning $G$-dependent probability distributions $p_G(\bfx)$. We now consider diffusion modeling of $p_G(\bfx)$ under a forward diffusion process $d\bfx = -\frac{1}{2}\bfH \bfx \, dt + d\bfw$ where $\bfH$ is symmetric positive semi-definite. 

Naively, one may choose $\bfH$ to be proportional to the identity (diffusing to an isotropic Gaussian), as is universally done for images and previously done for molecular conformer generation \citep{xu2021geodiff}. However, such a diffusion does not take into account the chemical graph structure and quickly disassembles the molecule into highly implausible states. Instead, we observe that a choice of $\bfH$ corresponds to a choice of an arbitrary Gaussian \emph{stationary distribution} of the diffusion:
\begin{equation}\label{eq:stationary}
    \lim_{t\rightarrow \infty} p(\bfx_t) \propto \exp\left(-\frac{1}{2}\bfx^T\bfH\bfx\right)
\end{equation}
We can re-interpret this distribution as the Boltzmann distribution $p(\bfx) \propto e^{-E(\bfx)}$ of an arbitrary \emph{quadratic potential} $E(\bfx) = \frac{1}{2}\bfx^T\bfH\bfx$. Similarly, we may re-interpret the forward diffusion as the Brownian motion of a particle under the same time-independent quadratic potential: $d\bfx = -\frac{1}{2}\nabla_\bfx E(\bfx) \, dt + d\bfw$. In the physics literature, such motion is known as \emph{overdamped Langevin} or \emph{Brownian dynamics} \citep{erban2014molecular} and is known to converge to the Boltzmann distribution of the potential, consistent with the formulation here.

This Brownian dynamics perspective on forward diffusions provides clear guidance on how to choose the drift term $\bfH$: we choose it so that undesired, chemically implausible structures have high energy $E(\bfx)$. Conceptually, this accomplishes two main objectives: (1) samples from the prior distribution are automatically consistent with the encoded chemical constraints; and (2) the forward (and later, reverse) diffusions maintain these constraints such that highly implausible structures are never reached. In \emph{harmonic diffusion}, we choose $E(\bfx)$ as the sum of quadratic or \emph{harmonic} constraints for each edge in $\mathcal{E}$, meaning:
\begin{equation}\label{eq:harmonic}
    E(\bfx) = \frac{\alpha}{2}\sum_{(i,j) \in \mathcal{E}}||\bfx_i - \bfx_j||^2
\end{equation}
Here, $\bfx_i, \bfx_j \in \mathbb{R}^3$ are the coordinates of the $i$th and $j$th nodes and $\alpha>0$ is interpreted as the strength of the edge. This potential is quadratic in the coordinates $\bfx$ and therefore can be written in the form $\frac{1}{2}\bfx^T\bfH\bfx$, where $\bfH$ depends on the graph. Intuitively, this potential and the consequent drift term $\bfH \bfx$ constrain adjacent nodes to be nearby in 3D space, resolving the most noticeable shortcoming (i.e., molecular disassembly) of previous isotropic diffusions.

For protein structures, one option is to define $\mathcal{V}$ to be the heavy atoms, $\mathcal{E}$ the set of bonds between them, and construct $\mathbf{H}$ based on a harmonic potential defined using those bonds. However, in this work, we focus on sampling residue-level protein structures, in which $\mathcal{V}$ is the set of residues and $\mathcal{E}$ the edges connecting neighboring residues. That is, a protein with $m$ residues is represented by a line graph $G$ of length $m$, in which $\bfx$ represents the coordinates of the alpha carbons. To construct the harmonic potential $\bfH$, we set $\alpha=3/3.8^2 \, \text{\AA}^{-2}$ to enforce a RMS distance of $3.8$ \AA\ between adjacent alpha carbons \citep{chakraborty2013protein}.

\subsection{Eigenmode Projections}

\begin{figure}
    \includegraphics[width=\textwidth]{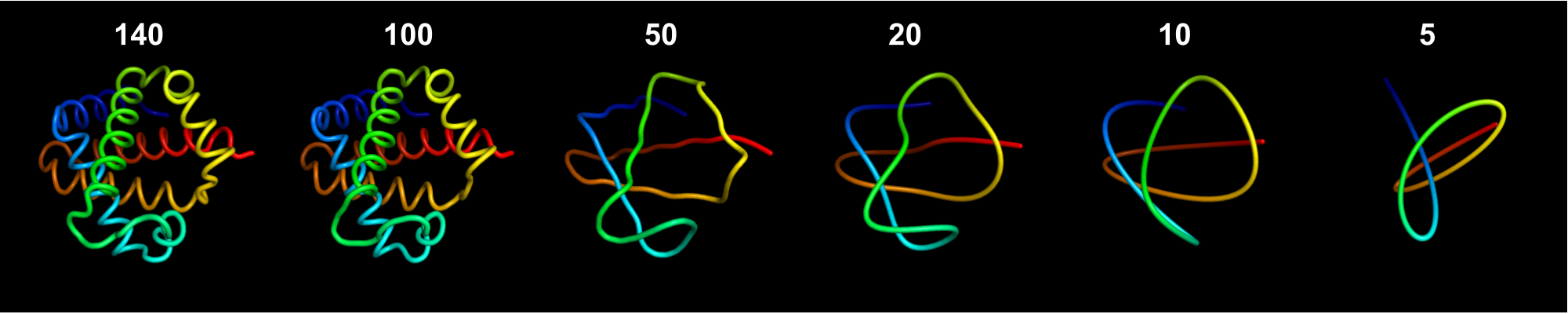}
    \caption{Eigenmode projections of the backbone structure of PDB ID \texttt{1a3n.A} onto progressively smaller numbers of modes. Due to symmetry, each mode is 3-fold degenerate. The full structure is described with 140 unique modes since the protein length is 141 (the last set of modes corresponds to translations). For ease of comparison, the structure is projected without perturbation; in the actual forward diffusion, noise is also injected into the remaining modes.}\label{fig:projections}
\end{figure}

The harmonic potential describes a forward SDE which can be used to train a score model and reversed via the Euler-Maruyama approach as described in \citet{song2021score}. However, the resulting reverse SDE is very stiff and requires a large number of reverse diffusion steps. To understand and solve this issue, we now study the behavior of this forward diffusion in more detail and propose an effective diffusion projection scheme.

Let $\bfH$ be decomposed as $\bfH = \bfP\bfLambda\bfP^T$ for orthogonal $\bfP$ and $\bfLambda = \diag(\lambda_1, \ldots \lambda_{3n})$ all nonnegative. Drawing an analogy with normal mode analysis in mechanics, we call eigenvectors of $\bfH$ (i.e. columns of $\bfP$) the \emph{normal modes} or \emph{eigenmodes} of the system, $\lambda_i$ the \emph{strength} of the modes, and $\bfy \equiv \bfP^T\bfx$ the coordinates along these modes. The diffusion kernel $p_{t\mid 0}$ and stationary distribution $p_\infty$ then both become uncorrelated (albeit nonspherical) Gaussians along the normal mode coordinates $\bfy$. The KL divergence between the perturbation kernel and the stationary distribution can then be decomposed as the sum of divergences in the coordinates along each mode:
\begin{equation}  \label{eq:kl}
    \KL(p_{t\mid 0} || p_\infty) = \sum_{i=1}^{3n}\left[ e^{-\lambda_it}\left(E_i - \frac{1}{2}\right) - \frac{1}{2}\log\left(1-e^{-\lambda_it}\right)\right]
\end{equation}
where $E_i \equiv \lambda_i (\bfy_0)_i^2/2$ is the initial energy in the $i$th mode.

The above expression fully describes the convergence of the forward diffusion towards the stationary distribution. In particular, the divergence along each mode decays with rate constant $\lambda_i$, which can vary by many orders of magnitude for different modes. Thus, the diffusion kernel will quickly converge to the stationary distribution along strong (i.e., large $\lambda_i$) modes, but will take much longer to converge along weak modes. This is analogous to the Born-Oppenheimer approximation in physics, where certain degrees of freedom equilibrate so rapidly that they are effectively stationary on time-scales relevant to other degrees of freedom. Indeed, we can characterize which degrees of freedom are “active” at any given diffusion time, starting with $3n$ at $t=0$ and reaching zero when we have converged to the joint stationary distribution. Nevertheless, regardless of the number of active modes, the stiffnes $\lambda_\text{max}/\lambda_\text{min}$ of the SDE remains very large, necessitating small step sizes for the entire duration of the sampled trajectory.

We now propose that, in both the forward and reverse diffusions, the structure is \emph{projected down to} only the modes that are still active at the given point in the diffusion (Figure~\ref{fig:projections}). That is, we set $(\bfy_t)_i=0$ for all modes $i$ such that $\lambda_it > \tau$ for some threshold $\tau$, thereby projecting the structure onto the subspace spanned by the remaining modes for which $\lambda_it \le \tau$. By construction, this reduces the stiffness $\lambda_\text{max}/\lambda_\text{min}$ of the SDE. At inference time, we start by sampling $(\bfy_T)_i \sim p_\infty$ from the stationary distribution only for the $k$ weakest (smallest eigenvalue) modes, where $k$ is a hyperparameter. Then, during the reverse diffusion, we successively sample $(\bfy_t)_j \sim p_\infty$ from the stationary distribution of the $j$th eigenmode just as it is about to become active, i.e., at $t_j = \tau/\lambda_j$. Similar to cascaded diffusion modeling of images \citep{ho2022cascaded}, this process induces a cascading-resolution generative process of the molecular structure, with the global structure being determined before local details (Figure~\ref{fig:overview}). Altogether, this procedure enables the sampling of large macromolecular structures with 100 or fewer Euler solver steps---significantly fewer than the 5000 steps required by \citet{xu2021geodiff} for much smaller molecules.

\subsection{Score Model Architecture}

Next, we develop a score model architecture $\bfs_\theta(\bfx, t)$ suitable for learning $\nabla_\bfx \log p(\bfx_t)$. We construct graph neural networks with tensor-product message-passing layers in \texttt{e3nn} \citep{thomas2018tensor, geiger2022e3nn}. The message-passing graph is constructed as a complete graph of size $m$, such that message-passing occurs between all pairs of residues. In addition to the residue coordinates, the score network is provided with node and edge features obtained by running OmegaFold on the input sequence and extracting the node and pair embeddings from the Geoformer stack. In this sense, our score model can be viewed as substituting for the deterministic structure module which usually operates on the Geoformer outputs. The score model is $SE(3)$-equivariant; since the stationary density (Equation~\ref{eq:harmonic}) is $SE(3)$-invariant, this ensures that the final model density will also be $SE(3)$-invariant \citep{xu2021geodiff}.

\subsection{Ranking Sampled Structures}

The trained \textsc{EigenFold} score model can sample multiple structures for a given protein sequence; however, it is often desirable to identify a single best structure prediction. For this purpose, we compute approximate lower bounds to model log-likelihoods for all sampled structures and select the one with the highest lower bound---i.e., the one most likely to have been sampled by \textsc{EigenFold}. Specifically, let $t_1, \ldots, t_K = T$ be a discretization of the forward and reverse SDEs. Then for any structure $\bfx$ (sampled or otherwise), we can compute a lower bound to the model log-likelihood as follows \citep{sohl2015deep}:
\begin{equation}    
    \log p(\bfx_0) \ge \mathbb{E}_{\bfx_{t_1\ldots t_K} \sim q}\log\left[p_\infty(\bfx_t) \prod_{k=1}^K \frac{p(\bfx_{t_{k-1}} \mid \bfx_{t_k})}{q(\bfx_{t_k} \mid \bfx_{t_{k-1}})}\right]
\end{equation}
Thus, by sampling a forward trajectory $\bfx_{t_1} \ldots \bfx_{t_K}$ starting from any given $\bfx_0$, we obtain a Monte-Carlo estimate of the evidence lower-bound (ELBO) for that structure. Notably, this insight removes the need to train a separate model to rank samples, as previously done by \citet{corso2022diffdock}.

\section{Experiments}
We train \textsc{EigenFold} on all structures deposited in the PDB on or before Apr 30, 2020 and validate on structures deposited between May 1, 2020 and Nov 30, 2020. To reduce training time, we train (and validate) only on structures with residue lengths between 20 and 256, for a total of 230,520 (14,128) training (validation) structures. To assess single-structure prediction accuracy, we make predictions for all CAMEO targets released between Aug 1, 2022 and Oct 31, 2022. After excluding targets with 750 or more residues for which OmegaFold embeddings could not be generated, the final test set consists of 183 CAMEO targets.

To assess the ability of \textsc{EigenFold} to model conformational diversity, we collect and filter two datasets from previous works. First, we collect 77 pairs of PDB IDs corresponding to fold-switching proteins from \citet{chakravarty2022alphafold2}. Second, we collect 90 pairs of apo/holo PDB IDs corresponding to ligand-induced conformational change from \citet{saldano2022impact}. For each dataset, we sample structures using the \texttt{SEQRES} entries of the PDB IDs with the shorter sequence or designated as "Apo," respectively. Both sets are filtered to remove pairs where the two sequences differ significantly in length or where the sequence used for sampling is 750 residues or longer.

\subsection{Single Structure Prediction}

\begin{figure}
    \centering
    \includegraphics[width=0.32\textwidth]{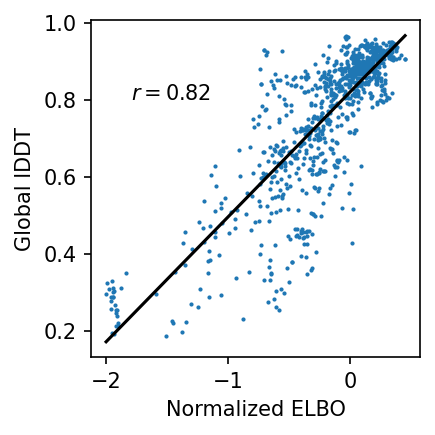}
    \includegraphics[width=0.32\textwidth]{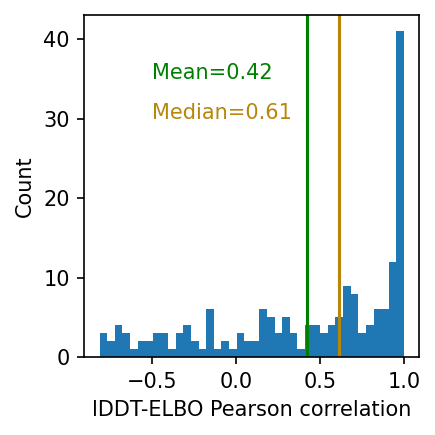}
    \includegraphics[width=0.32\textwidth]{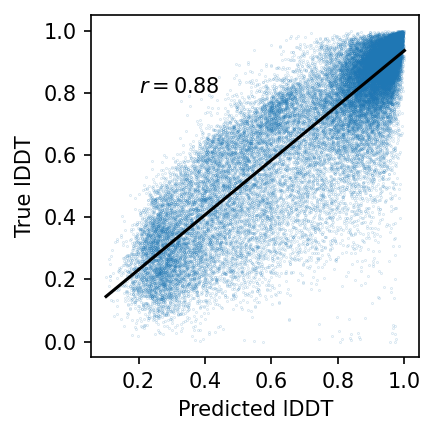}
    \caption{\emph{Left:} Scatterplot of normalized ELBO (i.e., divided by $(3n-1)$ where $n$ is the protein length) v.s. global lDDT. \emph{Center:} Histogram of per-target Pearson correlations between the normalized ELBO and lDDT. \emph{Right:} Scatterplot of predicted lDDT (average lDDT between sampled pairs) and true lDDT for all residues in the CAMEO test set.}\label{fig:elbo}
\end{figure}

For each of the CAMEO test targets, we sample five structures from \textsc{EigenFold} and compute the approximate ELBO for each. The top-ranked structure is considered the final prediction and is compared to the ground truth via standard metrics. Table~\ref{tab:cameo} compares the quality of these predictions relative to established methods RoseTTAFold, OmegaFold, ESMFold, and AlphaFold2. \textsc{EigenFold} samples are comparable in quality to those from RoseTTAFold, but fall short of the best results from AlphaFold2 and ESMFold. The approximate structural ELBO is well-correlated with absolute structural accuracy and thus serves as a good measure of model confidence (Figure~\ref{fig:elbo} (\emph{left}, \emph{center})). In particular, the positive per-target correlations (i.e., correlating only within the five samples for each target) on most targets justifies the use of the approximate ELBO as a means of ranking samples within a target.

\begin{table}
    \centering
    \caption{Single-structure prediction accuracy of \textsc{EigenFold} and baseline methods on CAMEO targets under 750 residues from Aug 1--Oct 31, 2022. All metrics are reported as mean / median.}
    \label{tab:cameo}
    \begin{tabular}{lcccc}
        \toprule 
        & RMSD$_{\text{C}\alpha}$ $\downarrow$ & TMScore $\uparrow$& GDT-TS $\uparrow$& lDDT$_{\text{C}\alpha}$ $\uparrow$\\ \midrule
        \textsc{AlphaFold2} & 3.30 / 1.64 & 0.87 / 0.95  & 0.86 / 0.91 & 0.90 / 0.93 \\
        \textsc{ESMFold} & 3.99 / 2.03 & 0.85 / 0.93 & 0.83 / 0.88 & 0.87 / 0.90\\
        \textsc{OmegaFold} & 5.26 / 2.62 & 0.80 / 0.89 & 0.77 / 0.84 & 0.83 / 0.89 \\
        \textsc{RoseTTAFold} & 5.72 / 3.17 & 0.77 / 0.84 & 0.71 / 0.75 & 0.79 / 0.82 \\
        \midrule
        \textsc{EigenFold} & 7.37 / 3.50 & 0.75 / 0.84 & 0.71 / 0.79 & 0.78 / 0.85\\
        \bottomrule
    \end{tabular}
\end{table}

Next, we find that the variability of the sampled ensemble is highly informative about the model error and can be interpreted as revealing model uncertainty. To measure this variability, we define, for any global measure of structural deviation $f$, an $f$-variation:
\begin{equation}
    f_\text{var} = \mathbb{E}_{\bfy_1, \bfy_2\sim \textsc{EigenFold}}\left[f(\bfy_1, \bfy_2)\right]
\end{equation}
where the expectation is approximated with the five samples. For example, if $f = \text{TM}$, then $\text{TM}_\text{var}$ measures the diversity of the sampled ensemble in terms of the average pairwise TMScore. We compare this quantity with $f_\text{exp}$, the average $f$ of the sampled structures relative to the ground truth. As illustrated in Table~\ref{tab:correls}, these measures are highly correlated across the CAMEO targets; thus, when the ground truth structure is unknown, the $f_\text{var}$ computed from sampled structures can be interpreted as a well-calibrated \emph{prediction} of $f_\text{exp}$ relative to the ground-truth.

The correlation between $f_\text{var}$ and $f_\text{exp}$ also holds for residue-level and pairwise accuracy metrics. In particular, we compute an expected lDDT for any given residue between pairs of sampled structures, and find that this is well-correlated with the lDDT for that residue between sampled structures and the ground truth (Figure~\ref{fig:elbo} (\emph{right}) and Table~\ref{tab:correls}). In this manner, we have access to a well-calibrated pLDDT for each residue, similar to the outputs of the confidence heads of existing structure prediction methods. Unlike existing methods, however, we can easily apply this framework to predict arbitrary error metrics without a bespoke confidence head. For example, in Table~\ref{tab:correls}, we illustrate that the aligned residue position error (i.e., error in residue position after RMSD alignment) and absolute pairwise distance error can be similarly predicted. Furthermore, the residue-level and pairwise metrics have high per-target correlations, indicating that they can be used to interpret the relative model confidence in different parts of the protein and their spatial relationships (Figure~\ref{fig:pair}).

\begin{table}
\begin{minipage}{0.5\linewidth}
    \caption{Pearson correlations between $f_\text{var}$ and $f_\text{exp}$ for various metrics $f$ of structural deviation. For residue-level or pairwise metrics, we compute a global correlation as well as a correlation for each target, reported as mean / median.}
    \label{tab:correls}
    \centering
    \begin{tabular}{lcc}
        \toprule
        & Global & Per-Target  \\
        \midrule
        \emph{Protein-level metrics} \\
        TM              & 0.88 & -- \\
        GDT-TS          & 0.90 & --\\
        RMSD$_{\text{C}\alpha}$            &  0.85 & -- \\
        lDDT$_{\text{C}\alpha}$  & 0.86 & -- \\
        \midrule
        \emph{Residue-level metrics} \\
        lDDT$_{\text{C}\alpha}$   &  0.88 & 0.73 / 0.81 \\
        Aligned position error & 0.80 & 0.68 / 0.75\\
        \midrule
        \emph{Pairwise metrics} \\
        Distance error & 0.75 & 0.69 / 0.72\\
        \bottomrule
    \end{tabular}
    \label{tab:my_label}
\end{minipage}\hfill
\begin{minipage}{0.47\linewidth}
    \centering
    \includegraphics{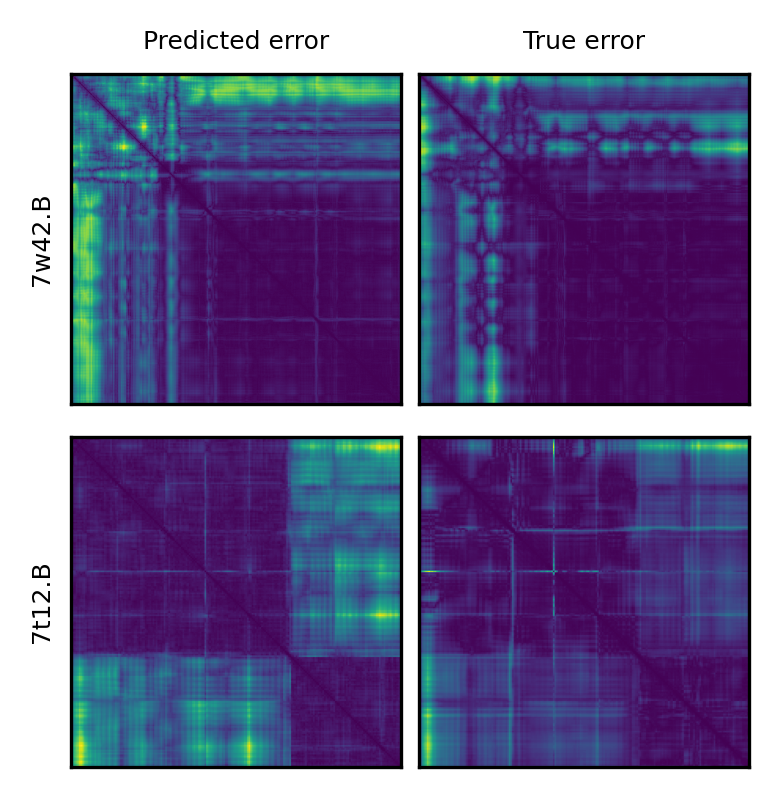}
    \vspace{-20pt}
    \captionsetup{type=figure}
    \caption{Predicted (left) and actual (right) pairwise distance errors for two CAMEO targets. Domains and inter-domain uncertainty are visibly and correctly predicted.}
    \label{fig:pair}
\end{minipage}
\end{table}

\subsection{Conformational Diversity}

To assess how well \textsc{EigenFold} can model protein conformational diversity, we sample five structures for each of the fold-switching and apo/holo pairs, and investigate the following questions:
\begin{enumerate}
    \item How well do the predicted structures model both conformations on a global level?
    \item Is the level of sample diversity predictive of the magnitude of the conformational change?
    \item Is the residue-level variation among samples predictive of true residue flexibility?
\end{enumerate}

\begin{figure}
    \centering
    \includegraphics[width=0.45\textwidth]{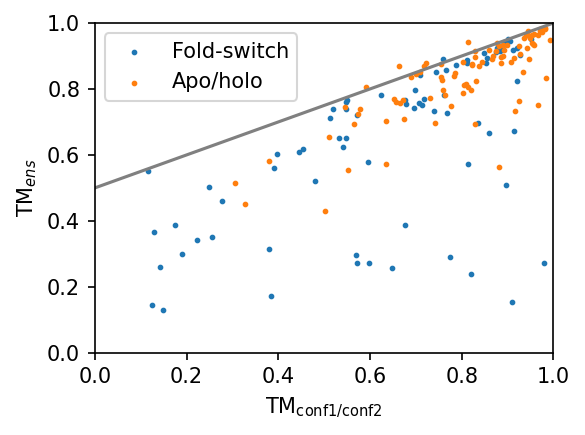}
    \includegraphics[width=0.45\textwidth]{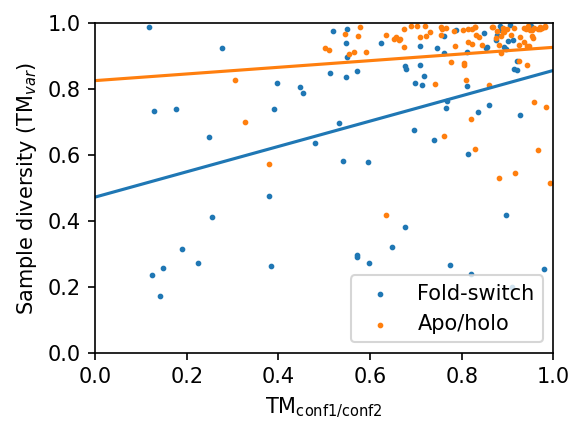}
    \caption{\emph{Left:} Coverage of the two ground truth conformations (TM$_\text{ens}$) plotted against the TM-score between the two conformations (i.e., the most dissimilar pairs are on the left). The gray line indicates a hypothetical baseline which always perfectly predicts one of the two conformations. \emph{Right:} Scatterplot of the TM-score between the two true conformations versus the average TM-score between sampled conformations.}
    \label{fig:diversity}
\end{figure}

To answer the first question, we define an ensemble TM-score as follows:
\begin{equation}
    \text{TM}_\text{ens}(\bfx_1, \bfx_2, \{\bfy_i\}) = \frac{1}{2}\left[\max_i \text{TM}(\bfy_i, \bfx_1) + \max_i\text{TM}(\bfy_i, \bfx_2)\right]
\end{equation}
where $\bfx_1,\bfx_2$ are the two ground truth structures and $\{\bfy_i\}$ are the \textsc{EigenFold} samples. This measures how well the sampled structures cover both ground-truth conformational states. Figure~\ref{fig:diversity} (\emph{left}) illustrates that \textsc{EigenFold} samples generally are a poor model of the two ground truth conformations, in the sense that they offer no improvement over a hypothetical baseline that always predicts a single conformation. Furthermore, the samples---even if different from each other---are generally very similar in terms of their deviation from the two ground truth structures, and tend to either heavily favor a single structure or model both structures relatively poorly.

\begin{table}
    \centering
    \caption{Pearson correlations between sample diversity and ground-truth diversity, measured in terms of TMScore (protein-level metric) or residue flexibility (i.e., absolute deviation after RMSD alignment). For the latter, we report global correlations and mean/median per-target correlations. }
    \label{tab:diversity}
    \begin{tabular}{lcc}
    \toprule
     &  Fold-switch & Apo/Holo\\
     \midrule
     TM & 0.36 & 0.12 \\ 
     Residue flexibility (global) & 0.23 & 0.13 \\
     Residue flexibility (per-target) & 0.28 / 0.26 & 0.41 / 0.40\\
     \bottomrule
    \end{tabular}
\end{table}

Next, to address the second and third questions, for each pair we compute the TM$_\text{var}$ (average pairwise TMScore) using the five sampled structures as a measure of sample diversity, and compare with the TMScore between the two ground truth conformations (TM$_\text{conf1/conf2}$). As shown in Figure~\ref{fig:diversity} (\emph{right}) and Table~\ref{tab:diversity}, TM$_\text{var}$ and TM$_\text{conf1/conf2}$ are moderately correlated. At the residue-level, we examine whether the flexibility of a residue within the sampled structures (i.e., average positional difference post-RMSD alignment) is predictive of the true flexibility of that residue under the conformational change. Table~\ref{tab:diversity} shows that both the global and per-target correlations are also positive but moderate. Altogether, while the conformational diversity and residue-level flexibility within sampled \textsc{EigenFold} structures are  somewhat informative of underlying conformational changes, the magnitude or residue-level localization of such changes are not modelled to high accuracy.

\section{Conclusion}

In this work, we developed \textsc{EigenFold}, the first diffusion generative model for predicting protein structures from a fixed protein sequence. In doing so, we built the first bridges between the rapidly advancing fields of diffusion modeling for molecules and modern structure prediction frameworks. Our model matches the performance of established methods on CAMEO targets and reveals model uncertainty via an ensemble of structural predictions, enabling customizeable ways to estimate and understand prediction error. We anticipate that these capabilities will prove important in downstream applications in which the relevant error could otherwise not be estimated using existing methods.

Although a generative modeling paradigm opens the door towards modeling conformational diversity and change, we find that \textsc{EigenFold} is currently unable to model these aspects of protein structure with high accuracy. Instead, it appears that the distribution of predicted structures is largely reflective of model uncertainty rather than underlying (i.e., aleatoric) uncertainty arising from flexibility. There may be several reasons for this gap: the model may not be accurate enough to resolve conformational changes of small magnitude; the training set consists largely of crystal structures that show little conformational flexibility; and the use of OmegaFold embeddings---without fine-tuning---may inject a bias towards the single-structure output of OmegaFold. Improving these aspects, with more tailored training settings or more expressive end-to-end score network architectures, could serve as promising directions of future work.

\section*{Acknowledgements}
We thank Hannes St\"{a}rk, Jason Yim, Jeremy Wohlwend, and Rohit Singh for helpful feedback and discussions. This work was supported by the NIH NIGMS under grant \#1R35GM141861 and a Department of Energy Computational Science Graduate Fellowship.

\bibliography{references}
\bibliographystyle{iclr2023_conference}

\end{document}